\documentstyle[epsf,twocolumn,aps]{revtex}
\begin{document}
% for two column  activate the line below...
\twocolumn[\hsize\textwidth\columnwidth\hsize\csname @twocolumnfalse\endcsname
\title{Structural Aspects of 
Magnetic Coupling in CaV$_4$O$_9$}
\author{W. E. Pickett}
\address{Complex Systems Theory Branch, Naval Research Laboratory,
Washington DC 20375}
\date{\today}
\maketitle
\begin{abstract}
The strong corrugation of the V$_4$O$_9$ layer in the
spin gap system CaV$_4$O$_9$ is 
examined for its impact on the exchange coupling constants between
the spin $\frac{1}{2}$ V ions.  Local spin density (LSD) calculations show
that the V spin occupies a $d_{x^2-y^2}$ orbital
($\hat x$ and $\hat y$ are the V-V directions).  The Kanamori-Goodenough
superexchange rules, and the degeneracy of ferromagnetic and
antiferromagnetic alignments within LSD, suggest that 2nd neighbor 
couplings dominate over
nearest neighbor, resulting in two coupled S=$\frac{1}{2}$
metaplaquette systems.
\end{abstract}
\pacs{PACS numbers: }

% for two column activate the line below...
]

%\section*{I. Introduction}
From its vanishing susceptibility $\chi(T\rightarrow 0)$\cite{taniguchi}
and related NMR properties\cite{ohama}
CaV$_4$O$_9$ is known to enter a quantum
disordered phase with a spin gap.  This behavior
has stimulated theoretical investigation
of the exchange couplings between the S=$\frac{1}{2}$ spins on the
V lattice,\cite{ueda,sano,troyer,staryk,albrecht,katoh,gelfand} 
using a Heisenberg model with 
nearest ($J_{nn}$) and next-nearest
($J_{nnn}$) exchange couplings.
The $\frac{1}{5}$-depleted lattice
\cite{bouloux} (described
below) has been viewed as an array of 
square ``plaquettes" of V ions tending toward singlet formation.
\cite{ueda}.  Isolated plaquettes have a singlet ground state; structural 
chemistry however suggests intra- and inter-plaquette
V-V coupling ($J$ and $J^{\prime}$) should be very 
similar, so the limit of isolated plaquettes is not realistic.

Depletion alone does not destroy N\'eel order,\cite{troyer} and although
competing interactions lead to $\chi$(T)$\rightarrow$0, they
do not account quantitatively for the $\chi$(T)
data.\cite{staryk,gelfand}  Yet the ``plaquette phase" of incipient
singlets provides an attractive framework for accounting for the 
lack of magnetic ordering and the spin gap behavior.  In this paper we
take into account all complications of the crystal structure, and find
that the orbital character of the occupied V $d$ spin orbital is
different than previously anticipated.
\cite{ueda,sano,katoh,marini}
For this spin orbital, nnn exchange coupling exceeds nn coupling.
The resulting coupling leads to a picture of two coupled
metaplaquette systems that may preserve the tendency toward local singlet
formation while enhancing frustration of ordering.

%Structure
The crystal structure of CaV$_4$O$_9$ \cite{bouloux}
has been idealized in most previous
theoretical treatments to the consideration of VO layers with
periodic V vacancies.  The actual structure is much more
interesting.  The space group is simple tetragonal P4/n (\#85 in
the International Tables), with $a=8.333 \AA$, $c=5.008 \AA$, and
two formula units per cell.  The low crystal symmetry is reflected in
the fact that ten of the 15 internal structural parameters (for five
sites) are not fixed by symmetry.  Conceptually, one may start from 
a VO square lattice with cations and anions
arranged as on a checkerboard.  Each V ion along a V-O
axis has an apical oxygen above (say) the layer, with V ions along
neighboring V-O lines having their apical oxygen below the plane. 
V ions are removed in a pattern corresponding to a left-moving
(or right-moving, giving the chiral partner) knight on a 
chessboard that lies at a $45^\circ$ angle derived
only from the V ions: from one V vacancy, go
two V ions along a line of V ions, and one to the left to locate a
neighboring vacancy.  The resulting pattern, shown in Fig. 1,
is a $\sqrt 5$ x
$\sqrt 5$ enlargement with respect to the V sublattice, which is
itself $\sqrt 2$ x $\sqrt 2$ larger than the VO square lattice.
Ca ions
arrange themselves either above or below the V vacancies.

% FIG. 1
\begin{figure}[tbp]
\epsfxsize=7cm\centerline{\epsffile{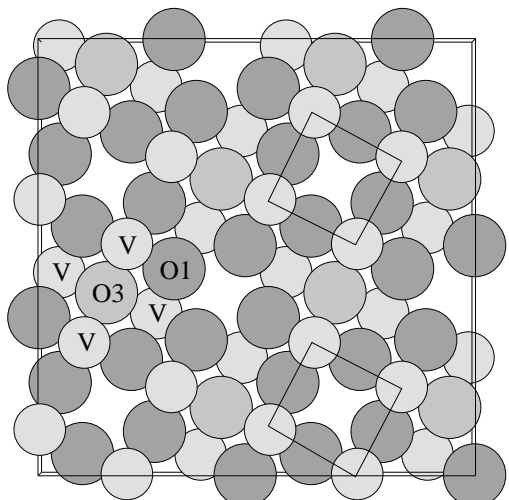}}
\caption{Top view of the V$_4$O$_9$ layer, with Ca and O2 atoms not
displayed.  A plaquette of V ions is labelled by `V'.  Two
metaplaquettes, formed by the V ions above the layer, are shown by
the solid lines.
\label{Fig1}}
\end{figure}

Superimposed on this superstructure is a severe corrugation 
of the plane, with V ions displaced alternately
$\pm$0.625 $\AA$ perpendicular to the plane in the
direction of their apical O ion.  The entire structure is shown in
Fig. 2.  V ions, as well as apical ions (O2),
remain equivalent by symmetry, but there are two other O sites.  O1 sites are
near ($\pm 0.12\AA$) the plane and are coordinated with three V ions,
two on one plaquette and the third on a
neighboring plaquette, all at 1.95-1.96 \AA.
O3 sites remain in the plane at
positions at the plaquette center, and are four-fold coordinated
with V ions at $2.04~\AA$.  The compound can be written descriptively
as [CaV$_4$O1$_4$O2$_4$O3]$_2$.  
The V, O1, and O2 sites have no symmetry.  
The V ion lies within a tilted
square pyramid of O ions (`square' of O1$_3$O3, with O2 apex), not far
from the center of mass of the five O ions (Fig. 3).  Each V$_4$
plaquette is boat-shaped, with two corner V ions up and the other two down. 
Due to the corrugation, the V-V-V angle that would be 180$^\circ$ for a
plane is reduced to 130$^\circ$. 

% FIG. 2
\begin{figure}[tbp]
\epsfxsize=7cm\centerline{\epsffile{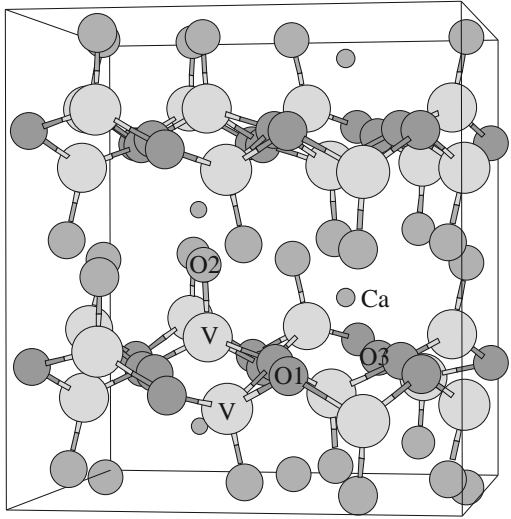}}
\caption{Structure of CaV$_4$O$_9$, from Bouloux and Galy,
%\cite{bouloux}
illustrating the severe
corrugation of the V-O layer.
Unit cells in two
successive layers are shown.
\label{Fig2}}
\end{figure}

% FIG. 3
\begin{figure}[tbp]
\epsfxsize=5cm\centerline{\epsffile{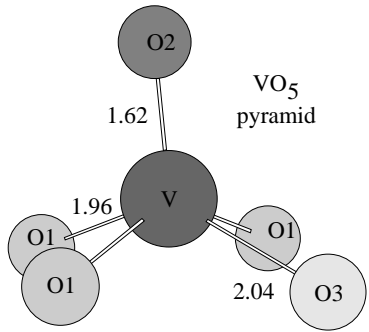}}
\caption{Side view of the VO$_5$ ``pyramid."  The V ion actually lies
above the center of mass of the five O ions.  V-O bond 
lengths ($\AA$) are shown.  The three V-O1 lengths are inequivalent
but nearly equal.
\label{Fig3}}
\end{figure}

%\section*{II. Method of Calculation}
We have applied the full potential linearized augmented plane wave (FLAPW)
method \cite{wei,singh} used in previous studies of
magnetic transition metal oxide compounds.
The sphere radii used in fixing the FLAPW basis were chosen to be
2.0, 2.0, 1.60, 1.45, and 1.60 a.u. for Ca, V, O1, O2, and O3,
respectively. 
Local orbitals were added to
the basis set for extra flexibility and
to allow semicore states to be treated within the 
same energy window as the band states.  For the V atoms, $s$, $p$,
and $d$ local orbitals were used, while for Ca and the O atoms $s$ 
and $p$ local orbitals were added.  The plane wave cutoff
corresponded to an energy of 19 Ry.  The total basis set size is
3350 LAPWs and local orbitals. 

An initial calculation without spin polarization found the O $2p$ states
to be centered 6-7 eV below the center of the V $3d$ bands, and 
separated from them by a gap of 2 eV.  The lowest Ca-derived band bands
arise from the $3d$ states, but lie above the V $d$ bands and are
inactive.
The Fermi level E$_F$ lies low in V $d$ bands,
corresponding to a $d^1$ configuration.  Thus the conventional ionic picture
is well respected in this compound.

To interrogate magnetic interactions, ferromagnetic (FM) and 
antiferromagnetic (AFM) alignments of the V moments were studied.
The energy gain from polarization is 0.12 eV/V ion.
For FM ordering, an insulating state
is obtained, as shown in Fig. 4, so the moment obtained is
precisely 1 $\mu_B$ per V.  An insulating result was not anticipated
from the non-polarized calculation: there was no gap in the $d$ bands,
so a rigid Stoner splitting
of majority and minority bands would leave a metallic result.
The lowest lying V $d$ orbital
in the majority bands, which we show below to be the $d_{x^2-y^2}$
orbital, upon polarization becomes separated from the remaining 4 $d$
orbitals, due to an exchange splitting $\Delta_{ex}$
that is strongly orbital dependent.
$\Delta_{ex}(d_{x^2-y^2})\approx$ 1.3 eV is an unusually large value
(LSDA exchange forces usually are less than
1 eV/$\mu_B$), reflecting a weakly hybridized (and therefore
more confined) $d_{x^2-y^2}$ orbital.  
$\Delta_{ex}$ is only
about 0.6 eV in the center of the $d$ bands and decreases to
0.4 eV at the top.  The calculated gap in the majority bands is
0.7 eV, and the gap between occupied majority and unoccupied 
minority bands is of the order of 0.1 eV.  Correlations effects can
only increase these gap values, perhaps substantially.  
%Since spin disorder is not likely to
%close the gap, this result already accounts
%qualitatively for the insulating character of the compound.

% FIG. 4
\begin{figure}[tbp]
\epsfxsize=5.5cm\centerline{\epsffile{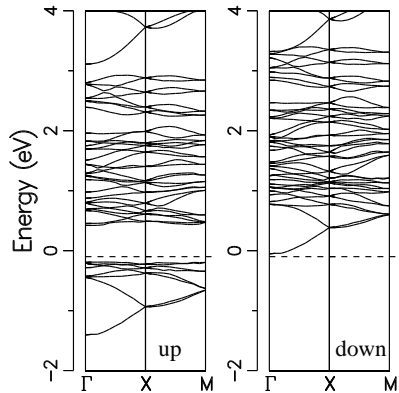}}
\caption{Majority (left) and minority (right) bands along high
symmetry directions for FM aligned CaV$_4$O$_9$.  The majority
$d_{x^2-y^2}$ is disconnected from other $d$ states, leading
to an insulating result. 
\label{Fig4}}
\end{figure}

The spin density for this FM state, shown in Fig. 5, is clearly that
due to occupation of a single spin-orbital of V $d_{x^2-y^2}$
character, where $\hat x$ and $\hat y$ refer to the lines of V
ions.  Due to the low V ion site symmetry the lobes do not point
precisely in the direction of neighboring V ions nor lie exactly in
the crystallographic $\hat x - \hat y$ plane, but these `misalignments'
are only a few degrees.  These
spin-orbitals are non-bonding with respect to the neighboring O ions,
a point we return to below.  This occupied orbital is not of the type
($d_{xy}, d_{yz}, d_{xz}$) anticipated in previous work,\cite{sano,katoh}
or used by Marini and Khomskii for orbital ordering 
($d_{xz}$ or $d_{yz}$).\cite{marini}
Note that this orientation is determined
by the crystal field, and is unrelated (at least for this narrow
band system) to the FM order.

The chosen AFM order was of the N\'eel type: each V spin is antiparallel
to its two neighbors on a plaquette and to the neighbor on the next plaquette.
This type of order breaks inversion symmetry, and all V ions on one
side of the (idealized V-O) plane have the same spin direction.  This ordered
state is essentially degenerate with the FM alignment.\cite{degen}  
The resulting
state is also insulating, and as expected the occupied bandwidth is smaller
(20\%) than for FM alignment.

%\pagebreak

% FIG. 5
\begin{figure}[tbp]
\epsfxsize=8cm\centerline{\epsffile{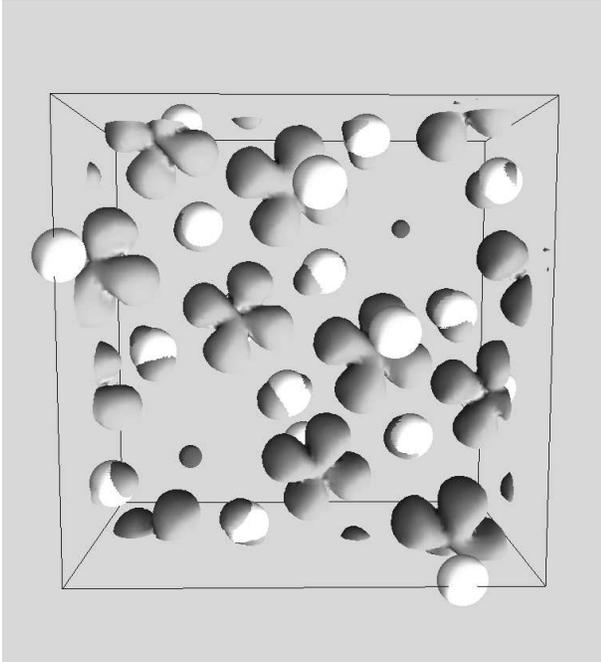}}
\caption{Top view of an isocontour of the spin density of FM 
aligned CaV$_4$O$_9$ (truncated at the cell 
boundaries), illustrating the
occupied $d_{x^2-y^2}$ V$^{4+}$ spin orbital.
The white spheres indicate oxygen sites, small dark spheres 
indicate Ca ions.
\label{Fig5}}
\end{figure}

The effective V $d_{x^2-y^2}-d_{x^2-y^2}$ transfer integrals $t_{ij}$
are thought to be mediated $via$ virtual hopping processes through the
intervening O ions, with amplitudes $t_{dp\sigma}$ and
$t_{dp\pi}$.
In the idealized VO parent layer the O ions lie at $45^\circ$
angles to the V-V nn direction.  In terms of hopping processes,
the nnn V ion is no more distant that the nn V ion and therefore
should have comparable hopping amplitude, hence
exchange coupling $J$.
According to the Goodenough-Kanamori (GK) rules,\cite{goodenough}
the hopping amplitudes depend strongly both on the symmetry of the
$d$ and $p$ orbitals involved and on the V-O-V angles.
The first point to note is that this spin orbital is orthogonal (except
for the $\phi \approx $ 5$^\circ$ rotation) to the O $p_{\sigma}$ 
orbital; hence the
coupling must proceed through the $p_{\pi}$ orbital. This coupling involves
only the $t_{dp\pi}$, which may be a few times smaller than $t_{dp\sigma}$.
The $p_{\pi}$ orbital couples to the $d_{x^2-y^2}$ orbital on the nnn V
(again via $t_{dp\pi}$).  The $p_{\pi}$ orbital however is {\it orthogonal}
to the $d_{x^2-y^2}$ orbital on the nn V (with respect to which it is
a $p_{\sigma}$ orbital).  Thus the usual second order nn V superexchange 
coupling vanishes for the idealized layer.  

Due to the symmetry lowering distortions, this vanishing is incomplete, 
and the surviving coupling may be larger than third order effects
involving polarization of the O ion.\cite{goodenough}  
$pd\sigma$ hopping is allowed as
$t_{dp\sigma} sin\phi$ and $pd\pi$ hopping is reduced to
$t_{dp\pi} cos\phi$. 
The couplings become
\begin{eqnarray}
J_{nn} &~ \propto &~2~t_{dp\pi}~cos\phi~t_{dp\sigma}~sin\phi/
(\varepsilon_d - \varepsilon_p), \\
J_{nnn}&~\propto &~t^2_{dp\pi}~cos^2\phi/(\varepsilon_d - \varepsilon_p), 
\end{eqnarray}
where the factor of two arises from the two paths, and
$\varepsilon_d - \varepsilon_p \approx$ 4-5 eV is the 
energy separation between V $d$
and O $p$ states.  Since $\sin\phi \approx$ 0.1, an rough estimation
gives $J_{nnn}/J_{nn}\approx \frac{1}{2}(t_{dp\pi}/t_{dp\sigma})\cot \phi
\approx \frac{1}{2}\cdot\frac{1}{3} \cdot 10 \approx 2$. (A rotation $\Omega$ 
necessary to rotate
the $d_{x^2-y^2}$ orbital into the plane of the V-O bond affects both
couplings proportionately, and has been neglected in this discussion.)

This already rough estimate will be altered by the structural corrugations.
nn V ions, which differ in $z$ coordinate by 1.25~\AA, are
connected along two V-O-V paths, whose contributions (presumably) add.
For nnn V ions, which are at the same height, there is a single path.
Within
a plaquette, the angles for nn V ions are 100$^\circ$ (V-O1-V) and 95$^\circ$
(V-O3-V), and is 144$^\circ$ (V-O3-V)
across the diagonal.  Between plaquettes, nn V are connected by
two identical 100$^\circ$ angles (V-O1-V), while the diagonal nnn angle
is 130$^\circ$ (V-O1-V).  These angular variations, coupled with the
fact that the O ions are not coplanar with the lobes of
the $d_{x^2-y^2}$ spin orbital and 
that direct V-V exchange may not be negligible,
make it very difficult to estimate realistic net exchange couplings.

The direct V-V separations are 3.00$\pm0.01 \AA$, which can
be compared to the
bondlength in V metal of 2.73 \AA.  This difference of less than
10\% suggests that direct V-V hopping might be
appreciable.  The less extended nature of the V$^{4+}$ orbitals compared 
to neutral V orbitals will reduce the overlap, however.
Since nn V ions do not lie in the plane of the $d_{x^2-y^2}$ orbital,
the orbital must be rotated by 
$\theta$=$\sin^{-1}(\frac
{1.25}{3.00})$=25$^\circ$ on both atoms before the overlap 
can be expressed in 
terms of the usual $t_{dd\sigma}, t_{dd\pi}, t_{dd\delta}$ hopping amplitudes.
Neglecting the latter two, which at this distance should be much 
smaller than $t_{dd\sigma}$, the effective hopping between the
non-aligned $d_{x^2-y^2}$ spin-orbitals will go as $t_{dd\sigma}
\cos^2 \theta$, a reduction by $\cos^2 \theta \approx$0.8.

It is not a straightforward matter to determine the 
relative importance of direct V-V hopping and V-O-V mixing
that gives the dispersion
pictured in Fig. 4.  The effect on the $d$ bandwidth of the V-O-V
coupling was probed by moving the O2 ions from
$\frac{z}{c}$=0.939 to $\frac{z}{c}$=0.75.
This distortion brings the O2 ion nearer one V but much farther from
the other (2.6 \AA), essentially eliminated V-O1-V coupling
of neighboring plaquettes.  The resulting 
occupied $d$ bandwidth of W$_d$=0.4 eV, compared to 1.3 eV 
for the real structure, can be ascribed to direct V-V coupling.  This
result suggests that coupling through the O1 ions is responsible for 
roughly $\frac{2}{3}$ of the bandwidth, with direct V-V interactions providing
the remaining $\frac{1}{3}$; {\it i.e.} direct V-V coupling is not negligible.
This complexity arises specifically from the fact that the spin-orbital
is $d_{x^2-y^2}$ character, which greatly hinders coupling to O ions
and maximizes V-V interaction.  Using a coordination number $z_c$=3,
the V-V bandwidth translates to a hopping amplitude $t_{dd\sigma}$
=W$_d$/(2$z_c \cos^2 \theta$)=80 meV.

The hybridization leading to magnetic coupling is reflected in the spin
density distribution on O ions neighboring  V.  The primary feature,
in both FM and AFM alignments studied, is an antiparallel polarization
of the apical O2 by perhaps as much as 0.2 $\mu_B$ (the spin density
attributable to a given ion is not precisely defined).  The O2 ion,
however, is not involved in exchange coupling.  The O1 
ions have moments about $\frac{1}{3}$ as large, aligned parallel, while
the O3 ion's moment is even smaller (zero by symmetry for AFM case).
Thus the O1 site should be more important in the exchange coupling process than
is the O3 site.

These results -- the occupied spin orbital of $d_{x^2-y^2}$
character, $J_{nnn} > J_{nn}$, $J_{V-V}\neq$0, degenerate FM
and AFM alignments -- suggest a picture of coupled two metaplaquette 
system.  Each metaplaquette is comprised solely of V ions either above the 
plane, or below the plane; the upper metaplaquette is highlighted in Fig. 1.
The two metaplaquette systems, each topologically equivalent to
the original, are coupled by at least three, 
possibly competing, 
exchange couplings $J_{nn}$, $J^{\prime}_{nn}$, and
$J_{V-V}$, but their values will not be easy to ascertain.  The
degeneracy of the FM and AFM alignments, in which the two 
ferromagnetic metaplaquette systems are aligned and antialigned
respectively, suggests relatively weak net coupling between the
two metaplaquette systems.

There is one particularly interesting difference between the original
plaquette system and (either one of) the metaplaquette systems.  As was
noted in the introduction, V-O-V couplings within a plaquette, or
between neighboring plaquettes, are very similar from the structural
chemistry point of view.  The metaplaquette system is different:
intra-metaplaquette coupling proceeds through the O1 site, while
inter-metaplaquette coupling proceeds through the O3 site.  Not only is the 
V-O3 distance 4\% longer than the V-O1 distance (Fig. 3)resulting
in weaker coupling, but the
environments are different and the LSD calculation indicates
substantially less hybridization through the O3 site, as reflected
in the much lower induced O3 spin density (somewhat visible in Fig. 3).  
This O1-O3 site distinction
provides a mechanism whereby the tendency to form singlets on each
metaplaquette is not disrupted by equally strong inter-metaplaquette coupling,
which was the case for the original plaquette model.

To summarize, it has been found that consideration of the full low
symmetry structure of CaV$_4$O$_9$ indicates that five unequal exchange
couplings ($J_{nn},J_{nnn},J^{\prime}_{nn},J^{\prime}_{nnn}$ from
superexchange, and direct $J_{V-V}$)
are present.  The orbital character of the spin, together with the
topology of the V-O layer, suggest that a coupled two metaplaquette
system may provide a realistic way to interpret the spin gap behavior.

%\section*{VII. Acknowledgments}
I thank R. R. P. Singh for stimulating my interest in this compound.
This work was supported in part by the Office of Naval Research.
Computation was done at the Arctic Region Supercomputing Center and
at the DoD Major Shared Resourse Center at NAVOCEANO.
 
% references


\begin{references}
\bibitem{taniguchi}S. Taniguchi {\it et al.}, J. Phys. Soc. Japan {\bf 64},
2758 (1995).
\bibitem{ohama}T. Ohama {\it et al.}, J. Phys. Soc. Japan {\bf 66},
23 (1997).
\bibitem{ueda}K. Ueda {\it et al.}, Phys. Rev. Lett. {\bf 76}, 1932 (1996).
\bibitem{sano}K. Sano and K. Takano, J. Phys. Soc. Jpn.
{\bf 65}, 46 (1996).
\bibitem{troyer}M. Troyer {\it et al.}, Phys. Rev. Lett. {\bf 76}, 3822 (1996).
\bibitem{staryk}O. A. Staryk {\it et al}., Phys. Rev. Lett. {\bf 77},
2558 (1996).
\bibitem{albrecht}M. Albrecht {\it et al.}, Phys. Rev. B {\bf 54}, 15856 (1996).
\bibitem{katoh}N. Katoh and M. Imada, J. Phys. Soc. Jpn. {\bf 64}, 4105 (1995).
\bibitem{gelfand}M. P. Gelfand {\it et al.}, Phys. Rev. Lett. {\bf 77}, 2794
(1996).
\bibitem{bouloux}J.-C. Bouloux and J. Galy, Acta Cryst. B
{\bf 29}, 1335 (1973).
\bibitem{marini}S. Marini and D. I. Khomskii, cond-mat/9703130.
\bibitem {wei} S. H. Wei and H. Krakauer, Phys. Rev. Lett. {\bf55}, 1200 (1985);
 D. J. Singh, Phys. Rev.
 B {\bf43}, 6388 (1991).
\bibitem {singh} D. J. Singh, {\it Planewaves, Pseudopotentials, and the
LAPW Method} (Kluwer Academic,
Boston, 1994).
\bibitem{degen}The AFM alignment had 1.5 meV/V ion higher energy than the
FM alignment, probably within the convergence of the calculation.
\bibitem{goodenough}J. B. Goodenough, {\it Magnetism and the Chemical
Bond} (Wiley, New York, 1963), Chap. IIIC.
\end{references}
\end{document}